\documentclass[%
 reprint,
 amsmath,amssymb,
 aps,
]{revtex4-2}

\usepackage{graphicx}
\usepackage{dcolumn}
\usepackage{bm}
\usepackage{color}

\begin{document}

\preprint{APS/123-QED}

\title{Manipulating single photon emitter radiative lifetime in transition-metal dichalcogenides through Förster resonance energy transfer to graphene}

\author{R.Eddhib$^1$} 
\author{S.Ayari$^1$}
\author{A.Hichri$^1$}
\author{S.Jaziri$^1$,$^2$}

\affiliation{$^1 $ Facult\'{e} des Sciences de Bizerte Laboratoire de Physique des Mat\'{e}riaux Structure et Propri\'{e}t\'{e}s , Universit\'{e} de Carthage,7021 Jarzouna, Tunisia\\
$^2$Facult\'{e} des Sciences de Tunis Laboratoire de Physique de la Mati\`{e}re Condens\'{e}e
D\'{e}partement de Physique, Université Tunis el Manar Campus Universitaire 2092 Tunis, Tunisia\\
}%

\date{\today}

\begin{abstract}
Structural defects can crucially impact the optical response of monolayer (ML) thick materials as they enable trapping sites for excitons. These trapped excitons appear in photoluminescence spectra as new emissions below the free bright exciton and it can be exploited for single photon emissions (SPE). In this work we outline criteria, within our frame work, by which single photon emission can be detected in two dimensional materials and we explore how these criteria can be fulfilled in atomically thin transition-metal dichalcogenides (TMD). In particular, we model the effect of defects, in accordance with the most common experimental realisations, on the spatial auto-correlation function of the random disorder potential. Moreover, we provide a way to control the radiative lifetime of these emissions by a hybride heterostructrue of the ML TMD with a graphene sheet and a dielectric spacer that enables the Förster resonance energy transfer process. Our work predict that the corresponding SPE's quenched radiative lifetime will be in the picosecond range, this time scale is in good agreement with the recently measured exciton lifetime in this heterostructures.
\end{abstract}

\maketitle


\section{\label{sec:level1}Introduction}

Transition metal dichalcogenides (TMDs) have become a hot spot for two-dimensional material investigations. In fact, atomically thin TMDs which follow the common formula MX$_{2}$ ( M=Mo, W and X=Se, S) have raised considerable attention in both physical science and material engineering due to their unique optical properties \cite{F1,F2,F3,F4,F5}. For example, owing to the spatial and dielectric confinement, the heavy electron and hole effective masses arising from the atomic d-orbitals, the optical spectra are governed by excitonic features, which are stable even at room temperature and exhibit anomalously large binding energies of about 0.5 eV \cite{F6,F7, F3,F8}. Moreover, the truly two dimensional nature of these materials offers a straight-forward integration into existing photonic chip technology and offers a rich playground, in manipulating their optoelectronic properties \cite{F6,F7,F3,F8,F9}.\\

More recently, these (ML)-thick crystals have also emerged as promising materials for quantum information processing where their optical response are characterised  by the presence of a new type of SPE provided by the defects \cite{F2,F10,F11}. A single-photon emitter is a light source, which emits one photon at a time into an individual photon mode \cite{F2,F10,F11}. In this context, single-photon generation, which is an important building blocks for quantum technology has been achieved with a variety of systems in the past, most notably with, semiconductor quantum dots \cite{F12}, atomic defects such as nitrogen–vacancy centres in diamond \cite{F13}, organic molecules \cite{F14}, parametric down conversion \cite{F15} and more recently,  two dimensional atomic crystals, such as hBN \cite{F16,F17,F18}. Each system has its advantages, but also its limitations.  In fact, the usability of single-photon sources critically depends on the stability of photon emission. It is, for example, still a challenge to find a single-photon source that is stable, can be replicated, and can be easily interfaced with electrical contacts — all desirable, if not essential, and features for efficient quantum communication devices\cite{F2,F10,F11} . For example, due to the  large surface-to-volume ratio, colloidal semiconductor quantum dot with diameters of a few nanometers  suffer from off times (blinking), spectral jumps (spectral diffusion), and even disappearance of photon emission (photobleaching) \cite{F2,F19,F20,F21,F22}. However, the results on TMD monalyers are promising on various fronts. These materials are inorganic, stable and it could be easily used to form an heterostructures with other two-dimensional materials, such as graphene \cite{F2,F10,F11}. \\
In fact, Many experimental group such as, Philipp Tonndorf et al \cite{F2}, have reported the observation of single quantum emitters in atomically thin $WSe_{2}$.  In their measurement they have not observed photo bleaching of the emitters for hours. Indeed, They prove that the same single center is still present after several days and cycling of the temperature between 300 and 10 K.  Moreover the spectral stability of the single-photon emitter is excellent, especially if excited near-resonantly. These observations make evidence that TMDs can be emerged as a novel platform in which to realize single-photon emission. This single-photon emission arises from excitons trapped in local potential wells created by structural defects in the ML and it occurs 50-120 meV underneath the free neutral exciton in $WSe_{2}$\cite{F2}.  The defect in the TMD ML will introduces confinement in all directions in these case atomically thin semiconductor $WSe_{2}$ can be considered as a host for quantum dot-like defects \cite{F2,F10,F11}.
These observations make evidence that TMDs can be emerged as a novel platform in which to realize stable single-photon emission. This is the key future of the TMD's materials, combining thin structure with stable SPE source, where other systems such as InGaAs-quantum dots (QD's) need to be buried about 100nm away from any surface in order to escape noise from charge traps appearing there \cite{F23,F24}. In the ML TMD case this surrounding material is not only absent, but can be designed purposefully by encapsulation the TMD by different materials. This proximity of TMD-QD's like to a surface allows potential for complete control of the emitter’s surroundings. As proof of concept demonstration we are going to study  the effect of coupling this TMD to a plasmonic structures (graphene\cite{F25}), which require nanometre-scale proximity to enable the Förster resonance energy transfer(FRET) mechanism \cite{F26,F27,F28}. This last process Manifest in the transfer of a photo excitation energy  from a donor material (TMD) to an acceptor material (graphene) \cite{F29,F30,F31} in other words its the convertion of an optical radiation emitted from a quantum device into so-called surface plasmon-polaritons on weakly doped graphene  \cite{F32,F33,F34}. We have chosen the weakly doped graphene as a quencher first of all because of its TMD matching structure  and secondly because it is affordable, nontoxic, photo-stable and environmentally friendly \cite{F35,F36}. Graphene derivative like graphene oxide finds extensive application in biosensors and chemo-sensors because of its properties, where it exhibits a remarkable quenching efficiency through FRET process where it reaches 97 $\% $. \cite{F37,F38,F39,F40,F41,F42,F43} we highlight that in our theoretical work the quenching efficiency have reached 99 $\%$ for very small separation distances \\
We shall state that in our work, we are not going to take into account variations in the trapping potential due to the atomic structure nor defect-induced changes in phonon characteristics. Our main focus lies on providing qualitative microscopic insights into the impact of localized bright exciton states on the optical response of ML TMD's. And by that in this article we are going to firstly state a dependence within our frame work that's going to allow us to determine the conditions for which we obtain a SPE and prouve it's existing, secondly we are going to introduce the novel quenching rate expression which is applicable for all TMD materials and we are going to study the effect of a variety of spacers with ranging thickness on the SPE radiative lifetime of an exemplary TMD ML ( $WSe_{2}$) with a weakly doped graphene sheet as a quencher material where we show the dependence of this radiative lifetime of the TMD's SPE's on this different spacer materials and its separation distance from the plasmonic material, then we are going to highlight the effect of the random disorder parameters on the quenched radiative lifetime for an exemplary heterostructure ($WSe_{2}$/SiO$_{2}$/graphene). Finally, we are going to briefly determine the quenched radiative lifetime in function of the interlayer distance for all the different TMD materials $MX_{2}$.

\
\section{\label{sec:level2}Defect as a single photon emitter (SPE) source}

The most trivial way to observe SPE is to ensure that, within a given detection spot size, only one quantum emitter exists. In this context, Photoluminescence excitation spectroscopy reveals the excitonic nature of the emitters and provides evidence that these single excitons can be originate from free excitons trapped in local potential wells created by structural defects in the ML. In an experimental device these defects will be detuned and that this could generate single-photon emission even for large numbers of defects. Most commonly, defects in TMDs and other two dimensional materials can be created by different mechanisms, like alpha-particle irradiation \cite{F44}, residual impurities \cite{F45}, atomic vacancy formation \cite{F46}, strain from the substrate, high temperature annealing, impurity doping\cite{F47} and chemical functionalization \cite{F45} . These localized sites are randomly distributed over each ML and it giving rise to a localization of excitons, which eventually causes stable and sharp emission lines, i.e., the single-photon emitters. \\

We model this structural imperfection (the inhomogeneous disorder that derived from defects), using a Gaussian random disorder potential that binds the center of mass of the electron-hole pair. This potential is created from a superposition of $N$ random plane waves with random direction $\theta_{i}$ , random phase $\phi_{i}$, and correlation length $L$.
\begin{align}
V (R) =\sqrt{\dfrac{2}{N}} V_{0}\sum_{i=1}^{N} \cos \bigg(\dfrac{2\pi}{L}X \cos \theta_{i}+ \dfrac{2\pi}{L} Y \sin \theta_{i}+ \phi_{i} \bigg)  
\end{align}
In the presence of defect, the momentum of the center-of-mass motion is no longer a good quantum number as in the case of free exciton. Such effects introduce an additional confinement in all directions similar to self-assembled ("natural") semiconductor quantum dots. However, the exciton in ML TMDs behaves as a massive particle subject to a disordered potential, leading to spatially localized eigenstates of the center of mass motion. 
To  carry out the calculations of the energy and the wave function of  localized exciton in disorder potential derived from defects  we need to resolve numerically the four-dimensional Schrodinger equations $H_{X} \Psi_{X}(\boldsymbol{\rho},\theta,\boldsymbol{R})=E_{X} \Psi_{X}(\boldsymbol{\rho},\theta,\boldsymbol{R})$ ( see Appendix A and ref\cite{F52} for more detail about the localized exciton hamiltonian $H_{X}$ and the different approximation taken in our numerical calcultation).\\

\subsection{\label{sec:level3}Effect of the correlation length on the localization-delocalization of energy states}

\begin{figure*}[ht!]%
\includegraphics*[width=0.85\textwidth,height=12cm]{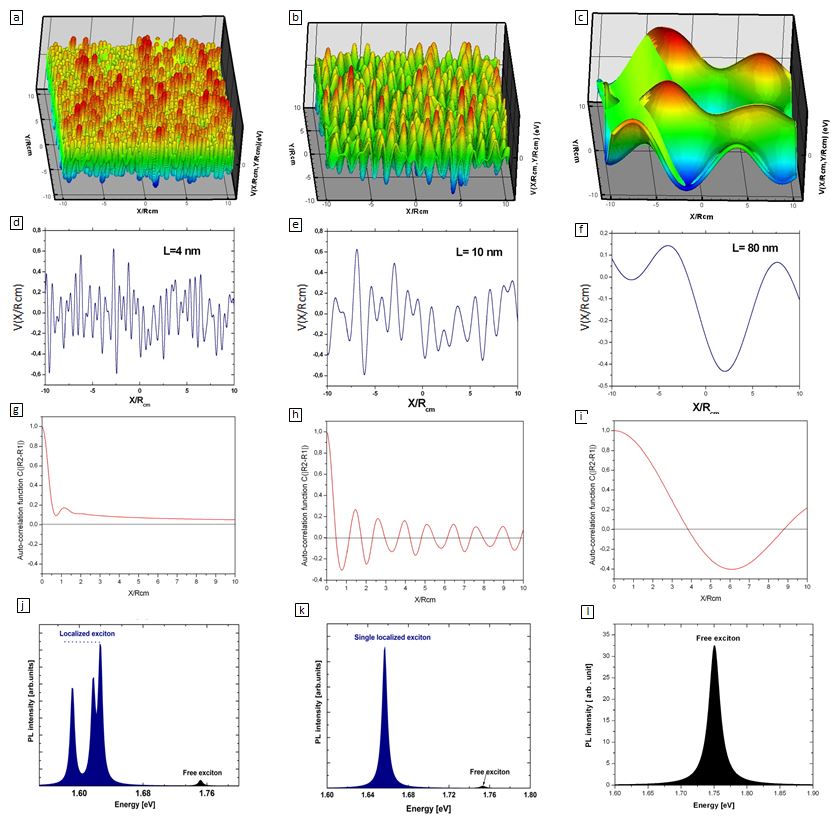}
\caption{(a-c) 3D-Schematic showing potential energy landscape due to defects under the influence of the correlation length,(d-f) a cut of the potential energy landscape for a fixed value of y (g-i) The auto-correlation function $ C $ is shown in function of $\vert R_{2}-R_{1} \vert$ where we highlight it's dependence on the correlation length. (j-l)PL spectrums of defects in ML WSe$_{2}$ for j) $L< R_{cm}$, k) $L\approx R_{cm}$ and k) $L> R_{cm}$ where we observe low-intensity PL peaks that are assigned to emissions from free excitons (pristine ML) and very intense peaks that are assigned to localized excitons we attribute this attenuation to the trapping of free exciton in the potential wells resulting in localized exciton we note that The  WSe$_{2}$ ML is deposited on the SiO$_{2}$/Si substrate and exposed to the air.}\label{pot}
\end{figure*}

Experimental and theoretical studies have proven that a disorder derived from defects can lead to a dramatic change in the physical behavior of the interband excitations, producing inhomogeneous spectral broadening and localization of all states, particularly in low dimensions system. In the same context, localization can also occur in time due to fluctuations \cite{F42}, even in the presence of interactions between particles, which is popularized in the recent publication with the term many-body localization \cite{F43}. We show in this part that for a random disorder potential and when its correlation length L is approximated to the average defect size, we have no localization of the exciton center of mass motion when the defect size is too small ( in order of 1 nm), while for big defect size  ( in order of 1 $\mu m$) the localization is maximized and it's impossible to obtain one SPE. It is important to note here, that we only obtain a SPE for an average defect size close to the centre of mass localization length magnitude ($\approx$10nm ). In this section, in order to assure the localization of one single state we are going to further investigate our theoretical model given in \cite{F52} to determine the auto-correlation function $C(\boldsymbol {R_{1}}, \boldsymbol {R_{2}})$ which determines the so-called scale of roughness of the potential surface through the correlation length L and by that we confirm the conditions for which we have a SPE. The spatial auto-correlation function of the disorder potential $V (\textbf{R})$ between tow position vectors of the  center of mass $\boldsymbol {R_{1}}$ and $\boldsymbol {R_{2}}$  can be written as \cite{F52}:

\begin{align}
C(\boldsymbol {R_{1}},\boldsymbol {R_{2}}) &=\dfrac{ \overline{\langle V(\boldsymbol {R_{1}})V(\boldsymbol {R_{2}})\rangle }- \overline{\langle V(\boldsymbol {R_{1}})\rangle}  \overline{\langle V(\boldsymbol {R_{2}})\rangle} }{ \overline{\langle V^{2}(\boldsymbol {R_{1}})\rangle}^{\frac{1}{2}} \overline{\langle V^{2}(\boldsymbol {R_{2}})\rangle}^{\frac{1}{2}}}  \notag \\ & = J_{0} \bigg(\dfrac{2\pi}{L}\vert \boldsymbol{R_{1}}-\boldsymbol{R_{2}}\vert \bigg)
\end{align}

here, $J_{0}(X)$ represents the Bessel function of the first kind order 0. The correlation function is quasi-periodic with respect to the lag $\vert \boldsymbol {R_{1}}- \boldsymbol {R_{2}}\vert $ as soon as $\vert \boldsymbol {R_{1}}- \boldsymbol {R_{2}}\vert\geq L$, with a pseudo-period roughly equal to \textit{L}. Notably, for $\boldsymbol {R} \in \mathbb{R}^{2}$ our disorder potential is characterized by zero-mean  $\overline{\langle V(\textbf{R})\rangle} =0$ and constant standard deviation equal to $\sigma_{V}=\overline{\langle V^{2}(\textbf{R})\rangle}^{\frac{1}{2}}=V_{0}$. It is worth noting that we can also define the  auto-covariance function which is just the unnormalized correlation function $C_{0}(\boldsymbol {R_{1}},\boldsymbol {R_{2}}) =\sigma_{V}^{2}.C(\boldsymbol {R_{1}},\boldsymbol {R_{2}})$.  Note that $C_{0}(\boldsymbol {R_{1}},\boldsymbol {R_{2}})$  tend to $\sigma_{V}^{2}$ when $\vert \boldsymbol {R_{1}}-\boldsymbol {R_{2}}\vert \rightarrow 0$. In the following we only address the property of the auto-correlation function, an important feature for this function is the decay shape which represents the roughness degree of the random surfaces (for a truly random rough surface, $ C(\boldsymbol {R_{1}},\boldsymbol {R_{2}}) $ usually decays to zero with the increase of $\vert \boldsymbol {R_{1}}- \boldsymbol {R_{2}}\vert$, where for a not really rough surface the auto-correlation function fluctuates around zero with the increase of $\vert \boldsymbol {R_{1}}- \boldsymbol {R_{2}}\vert$, and the larger the fluctuations are the less the roughness is ) while the spatial decay rate depends on the distance over which two points on this potential surface becomes uncorrelated and this represents the correlation length L of an auto-correlation function \cite{F53}.

In order to investigate the effect of the correlation length $L$ on the roughness of the random surface and hence its effect on the localization of the exciton center of mass motion, we plot in Fig.1 the disorder potential landscapes ( 3D and 1D) as well as the auto correlation function for three different value of L. We can clearly notice that the correlation length strongly affect the shape of the autocorrelation function as well as the potential roughness, which control of  localization / delocalization of the exciton states.  In fact, the roughness of the potential surface increases with  L. Therefore , based on the center of mass localization length $R_{cm}= \sqrt{\frac{\hbar}{M_{X}\omega_{cm}}}$, three regime can be distinguishable: i) $L < R_{cm}$  (fig \ref{pot}.(a,d,g)) , this correspond to strong roughness  of the  potential surface and lead to unbound exciton center of mass states in a random potential, which can be explained by the expansion of its wave function over the  heights of the potential lowering the role of potential fluctuation. This delocalized states regime is confirmed by the auto-correlation function where $C(\boldsymbol {R_{1}},\boldsymbol {R_{2}})$ decreases monotonically from 1 to  0 with the increase  $\vert\boldsymbol {R_{1}}- \boldsymbol {R_{2}}\vert$.
ii) $L> R_{cm}$, fig \ref{pot}.(c,f,i), this gives rise to a large potential width which lead to strong disorder. In this case, the potential surface is smoother than the other regime ( see fig 1.(i)) , and the exciton center of mass motion is strongly localized in the random disorder potential. In fact , in this regime since the localization length is very small compared to the correlation length, the confinement energy is large, so the quantification of the center of the exciton mass motion into only one state is not likely to occur at these potential wells and we can't obtain a SPE source in this regime. Hence,  in order to assure the presence of a SPE we shall define the third regime iii) where L is of the order of $R_{cm}$ we obtain one single localized state in the disorder potential. We interpret this by the fact that defect size is approximated to the correlation length. Taken ML WSe$_{2}$ as an exemplary TMD, these results are clearly shown in Fig.\ref{pot}(j-l), in which  we calculate  the PL spectrum at T = 4 K, using the following expression:\\

$P^{X}\propto \Big\vert \phi_{\tilde{1s}} (\rho=0) \Big\vert^{2}  \Big\vert \int \psi^{CM}_{n_{x},n_{y}}\boldsymbol{(R)} \, d^{2}\boldsymbol{R} \Big\vert^{2} \sigma(\hbar\omega-\hbar \omega_{X})$, where the lorentzian $ \sigma (\hbar \omega-\hbar \omega_{X})=\frac{\sigma_{X}}{\pi[(\hbar\omega-\hbar \omega_{X})^{2}+\sigma_{X}^{2}]}$  express the energy conservation taking into account the state broadening. Here, $\omega_{X}$ is is the optical transition frequency with $E_{X}=\hbar \omega_{X}$, the phenomenological parameter $\sigma_{X}$ is the half width at half maximum of the lines. $\psi^{CM}_{n_{x},n_{y}}(R)$ is the center of mass wave function of localized exciton, $\phi_{\tilde{1s}} (\rho)$ is the relative  wave function of the exciton. For ($ L >R_{cm}$) , i.e  larger potential, the calculated spectra display several anharmonic peaks, which simulate in a realistic manner the low-energy peaks observed experimentally, while for $ L \simeq R_{cm}$ and deep potential,  we found a single localized state which is also in an argument with the experience and the last one for $L< R_{cm}$ where the centre of mass motion is free . In the follow we will concentrate our study on ML WSe$_{2}$ where $ L= 100\r{A}$ and $ R_{cm}=80\r{A}$  in which we have a SPE that confirms what found in the experimental work \cite{F2}. After obtaining the SPE source we shall now present the way to control this TMD QD like system radiative lifetime by allowing the Förster resonance energy transfer process to occur  when the ML TMD is coupled with the doped graphene sheet.

\section{Förster resonance energy transfer (FRET) : Theoretical approach}
\begin{figure}[ht]%
\begin{center}
\includegraphics[width=0.5\textwidth,height=4cm]{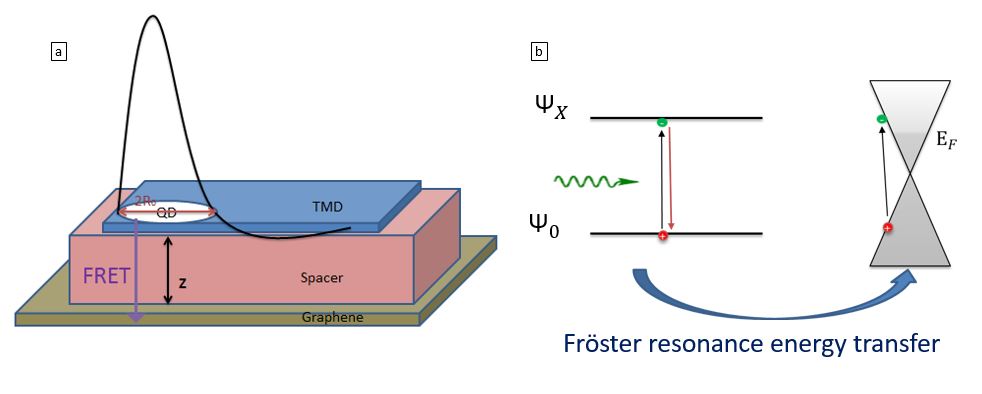}
 \caption{(a)Schematic representation of the hybrid heterostructure of the FRET system, (b)The Förster resonance energy transfer mechanism from the TMD quantum dot like system to the graphene sheet.}\label{fig:0}
 \end{center}
 \end{figure}

In this part we are going to introduce the FRET quenching rate expression which we have derived it from Kirill A. Velizhanin et al. \cite{F32} calculations in a way that fits our heterostructures. The procedure to make this FRET allowing structures consist of growing a dielectric spacer on top of a graphene sheet with a controllable thickness, then depositing a $TMD$ ML on top of the dielectric layer. The schematic hybrid system and FRET prosses is presented in  Figure.\ref{fig:0}.
 
The energy transfer from the TMD ML to the graphene sheet is based on the Coulomb interaction between the quasi-particle plasmon in graphene and the exciton of TMD QD like system, corresponding to the Förster coupling. We define this interaction by the following expression:
\begin{equation}\label{eq:1}
H_{int}=  \int d\textbf{r} V_{dip}(\textbf{r}) (\vert\Psi_{X}\rangle \langle \Psi_{0}\vert +\vert\Psi_{0}\rangle \langle \Psi_{X}\vert)\hat{\rho} (\textbf{r})
\end{equation}
where $\hat{\rho} (\textbf{r})=-e\chi(\textbf{r}) \chi^{*}(\textbf{r}) $ is the the charge density of graphene, with $\chi(\textbf{r}) $ and $\chi^{*}(\textbf{r})$ are the created and destroyed operator of an electron at position $\textbf{r}$ within the graphene ML. The absolute value of the electron charge is denoted by $ e = |e|$. $\vert \Psi_{X}\rangle$ and $ \vert \Psi_{0}\rangle$ denotes the localized exciton and the vacuum state.  The theoretical investigation of the FRET rate is done by evaluating the quenching rate that's expressed using the Fermi’s Golden rule and it results in \cite{F32} :
\begin{equation}\label{eq:2}
\Gamma_{q}= -\dfrac{2}{(2 \pi)^{2} \hbar} \int d\textbf{q} |V_{dip}(\textbf{q})|^{2} Im[\Pi^{rpa}(\textbf{q} ,\hbar \omega)]
\end{equation} 
We have been using the polarization operator $ \Pi^{rpa} $ of the graphene that we have calculated with in the Random Phase Approximation (RPA) because the polarization operator within the bare bubble approximation $ \Pi_{0}(\textbf{q} ,\hbar \omega) $ does not include the graphene’s polarization selfconsistently which can become crucial at nonzero doping levels ($\mu > 0$), where the finite carrier density at the Fermi level leads to the efficient Coulomb screening within the graphene sheet\cite{F33}. Here $\mu$  is the chemical potential of the graphene sheet. So we evaluate the polarization operator within (RPA) as $\Pi^{rpa}(\textbf{q} ,\hbar \omega)=\dfrac{\Pi_{0}(\textbf{q} ,\hbar \omega)}{1-V(\textbf{q} )\Pi_{0}(\textbf{q} ,\hbar \omega)}$ \cite{F34} where $ V(\textbf{q} ) = 2 \pi e^{2}/ \tilde{k}q $ is the two-dimensional Fourier transform of the Coulomb potential within the graphene plane and $\tilde{k}=\frac{\varepsilon_{spacer}+\varepsilon_{vac}}{2}$ donates the effective dielectric constant encapsulating the graphene sheet. 
In this work we are going to consider that $D < z$ where z is the  interlayer distance (spacer thickness) between the TMD QD-like and the graphene, and D is the defect potential diameter. In the appendix B we detail the calculation of $V_{dip}$  and it results (Eqs. (B12) ) in the next expression of the quenching rate:

\begin{equation}\label{eq:3}
\Gamma_{q}=-\dfrac{2 \pi e^{2}}{\hbar \, \varepsilon^{2}_{spacer}} ( 2d_{z}^{2}+d_{\parallel}^{2})\int_{0}^{\infty}d\textbf{q} \,  \textbf{q}  Im[\Pi^{rpa}(\textbf{q} ,\hbar \omega)]e^{-2qz}
\end{equation}
where $d_{\parallel}$ and $d_{z}$ are the projections of the ML TMD transition dipole d onto the graphene plane and the normal to this plane,respectively. We average these quantities over all the possible orientations with respect to the graphene plane which yields in $\langle  d_{z}\rangle$ and $\langle d_{\parallel}\rangle$.
Here,  $\langle  d_{z}\rangle=e.a_{0}$  where $a_{0}$ is the thickness of TMD ML, or $\langle d_{\parallel}\rangle$  can be expressed as a function of optical dipole transitions by electromagnetic waves perpendicular to the layer, and it's evaluated through elements of the optical matrix which are related to the rate of spontaneous recombination of the exciton:

\begin{align}
\langle d_{\parallel}\rangle^{2}& =\dfrac{ e^{2}}{m_{0}^{2}\omega_{X}^{2}}  \bigg\vert \phi_{\tilde{nl}} (\boldsymbol{\rho=0}) \bigg\vert^{2} \bigg \vert \int \psi^{CM}_{n_{x},n_{y}}(\boldsymbol{R}) d^{2} \boldsymbol{R} \bigg\vert^{2} \notag \\ & \bigg \vert \bigg \langle u_{c}\vert \boldsymbol{\epsilon}_{\boldsymbol{q}}^{\lambda}.\hat{\boldsymbol{p}} \big \vert u_{v} \bigg\rangle \bigg\vert^{2}     
\end{align}
here $\boldsymbol{\epsilon}_{\boldsymbol{q}}^{\lambda}$ is a unit vector characterizing the optical mode polarization $\lambda$,$\textbf{q}$ is the photon wave vector and $\hat{\boldsymbol{p}}$ is  the electron momentum operator. $\int \psi^{CM}_{n_{x},n_{y}}(\boldsymbol{R}) d^{2}\boldsymbol{R}$ represents the Fourier transform of the center of mass wave function taken at $\boldsymbol {K}=0$. The momentum matrix element $\langle u_{c}\vert \boldsymbol{\epsilon}_{\boldsymbol{q}}^{\lambda}.\hat{\boldsymbol{p}} \big \vert u_{v}\rangle$  and the exciton  wave function give rise to different selection rules. For the exciton center of mass  wave function only the wave function with even $n_{x}$, $n_{y}$ quantum number, giving a non-zero contribution in the calculation of the dipole moment. Also, $d_{\parallel} \neq0$  only for $\tilde{1s}$ states in which the angular momentum l=0. The selection rules come also from  $\langle u_{c}\vert \hat{\boldsymbol{p}} \big \vert u_{v}\rangle$  which is depend on the nature of the Bloch functions. For bright exciton emission, the only non-zero elements of the valence-conduction coupling terms are $\langle u _{c}\vert \hat{p_{\pm}} \vert u _{v}\rangle$ for circularly polarized light $\sigma \pm $ propagating along the normal to the sample ($ p_{\pm}= \dfrac{p_{x}\pm p_{y}}{\sqrt{2}}$), so that only optical modes with in-plane polarization components couple to these excitons. Therefore, using the $\boldsymbol{k.p}$ two-band model approximation, the momentum matrix element  is given by :
\begin{align}
\bigg \langle u_{c}\vert \boldsymbol{\varepsilon_{\pm}}.\hat{\boldsymbol{p_{\pm}}}  \vert u_{v} \bigg \rangle = \sqrt{\dfrac{m_{0}E_{p}}{2}}.
\end{align}
with $E_{p}=\dfrac{m_{0}E_{g}}{m_{e}}$ is the Kane energy for TMD materials  obtained in the two band mode. $E_{g}$ is the band gap energy. and $m_{e}$ is the electron mass.Now, inserting the Eq(6), and (7) in the Eq.5  the quenching rate can be rewriten as follow:
\begin{eqnarray}\label{eq:7}
\Gamma_{q}&&=-\dfrac{3 \pi e^{2}c^{3}}{4 n_{0}\omega_{X}^{3} \, \varepsilon^{2}_{spacer}} \bigg( \dfrac{2(e \,. a_{0})^{2}}{\langle d_{\parallel}\rangle^{2}}+1\bigg)\Gamma_{TMD} \nonumber\\ && \notag \times \int_{0}^{\infty}d\boldsymbol{q} \,\boldsymbol{q} Im[\Pi^{rpa}(\boldsymbol{q},\hbar \omega)]e^{-2qz}. \nonumber\\ 
&&=\int_{0}^{\infty} d\boldsymbol{q} \, F(\boldsymbol{q},\hbar \omega).
\end{eqnarray}
with $\Gamma_{TMD}$ is the spontaneous emission rate of localized exciton state j which is given by  :
\begin{eqnarray}
\Gamma_{TMD} && =\dfrac{4 e^{2}}{3\hbar m_{0} c^{3}}  n_{0}\omega_{X} E_{p} \bigg\vert \phi_{\tilde{nl}} (\rho=0) \bigg\vert^{2} \bigg \vert \int \psi^{CM}_{n_{x},n_{y}}(\boldsymbol{R}) d^{2}\boldsymbol{R} \bigg\vert^{2}\nonumber\\
&& =\dfrac{8 n_{0} \omega_{X}^{3}}{3 \hbar c^{3}}.\langle d_{\parallel}\rangle .
\end{eqnarray}
here $n_{0}=\sqrt{\tilde{k}}$ is the effective optical refraction index of the crystal environment.  Using the above relations we can now calculate the quenched radiative lifetime $\tau$ which is given by  $\tau^{-1}(z)=  \Gamma_{q}+\Gamma_{TMD}$ :
\begin{eqnarray}\label{eq:8}
\tau^{-1}(z)=&&\Gamma_{TMD}  \bigg[1- \dfrac{3 \pi e^{2}c^{3}}{4 n_{0}\omega_{X}^{3} \, \varepsilon^{2}_{spacer}} \Bigg( \dfrac{2(e.a_{0})^{2}}{\langle d_{\parallel}\rangle^{2}}+1\bigg)\nonumber \\ && \times \int_{0}^{\infty}d\boldsymbol{q} \, \boldsymbol{q} Im[\Pi^{rpa}(\boldsymbol{q},\hbar \omega)]e^{-2qz}\Bigg].
\end{eqnarray}
We can also define the quenching efficiency \cite{F50} :
 
 \begin{equation}
 \varphi_{eff}(z)=\dfrac{ \Gamma_{q}}{ \Gamma_{q}+\Gamma_{TMD}}=\dfrac{\Gamma_{q}}{\tau^{-1}(z)}
 \end{equation}
Our result coincides nicely with the result given in Ref \cite{F33}. Note that we obtain a different expression compared to it since we are interested in the TMD's behaviour and focused on it's excitons radiative decay rates. 
Using the last equations we gain the seeked way to control the SPE radiative lifetime.

\subsection{ Investigation of the graphene effect on the TMD's SPE's }

 \begin{figure}[ht]
 \begin{center}
 \includegraphics[scale=0.4]{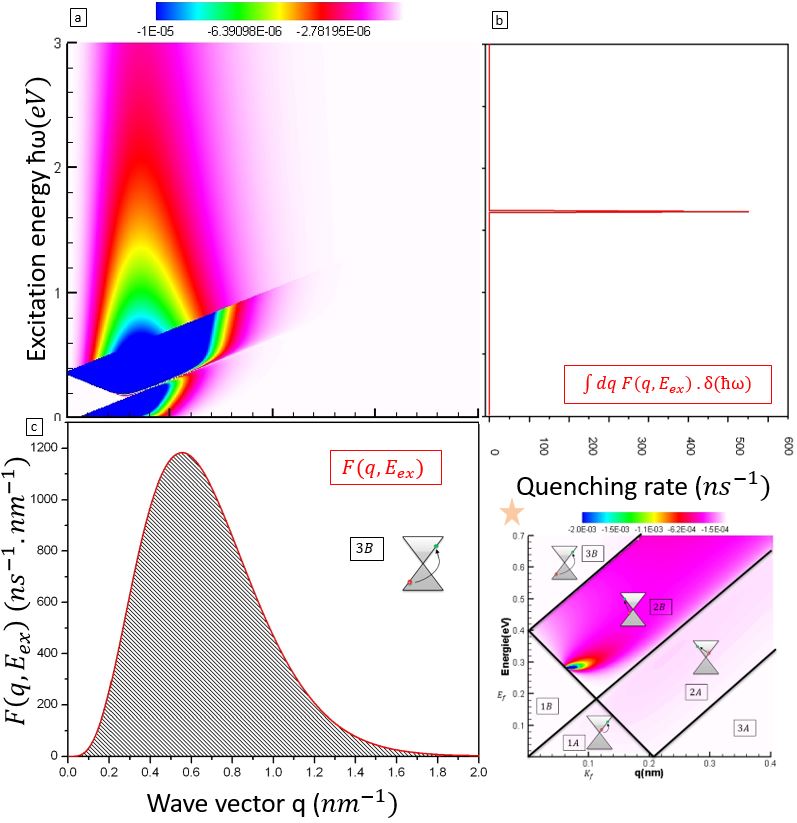}
 \caption{(a) Plot of the $Im[\Pi^{rpa}(q, \hbar \omega)]e^{-2qz}$ for a fixed distance between the TMD and the graphene layer (Z=5nm). ($\star$) represents the the Density plot for (z=0 nm), the white coulour emphasizes Region 1B $(q < \omega / v_{f}$, $q < 2q_{f} - \omega / v_{f}$) of the phase space and it is protected from Landau damping arising from both interband and intraband transitions. The coloured regions are 1A ($\omega / v_{f} < q < 2q_{f} - \omega / v_{f}$) and 2A ($\omega / v_{f} < q < 2q_{f} + \omega / v_{f}$, $q > 2q_{f} - \omega $), dominated by Landau damping resulting from intraband transitions, and 2B ($2q_{f} - \omega / v_{f} < q < \omega / v_{f}$, $\omega / v_{f} < 2q_{f}$ + q) and 3B ($q < \omega / v_{f} - 2q_{f}$) dominated by indirect and direct interband transitions, respectively.\cite{F33}. (b) represents the quenching rate of the SPE for an interlayer distance z=5 nm and for a fixed excitation energy $E_{X}=1.654$ eV. (c) the red line represents the shape of the density-density Imaginary part operator in function of wave vector q and for the same fixed values where the hatched part represents the quenching rate in function of the wave vector q. } \label{fig:1}
 \end{center}
 \end{figure}
 we exploit the derived equations to calculate the Förster induced quenching rate, efficiency  and radiative lifetime of the SPE, for an exemplary TMD donor material, ML tungsten diselenure ($WSe_{2}$) in a hybrid heterostructure with a graphene sheet as a quencher material and $SiO_{2}$ as a spacer. We are going to adopt the data Provided by the first part where we prove that for a given L, $V_{0}$, $R_{cm}$ and N we have an SPE source with an energy of 1.654 eV and a decay rate of $\Gamma_{TMD}=0.625 ns^{-1}$. These disorder parameters will be the 1st key in enabling the FRET process. We begin our work by exploiting Eq.(\ref{eq:7}), where we show In Fig.\ref{fig:1}-(a), a plot of the imaginary part of the polarization operator of the weakly doped graphene layer multiplied by the distance quenching factor $e^{-2qz}$  as a function of the excitation energy $\hbar \omega$ (eV) and the wave vector q ($nm^{-1}$) where we consider that the Fermi energy $E_{F}=0.2eV$ and the Fermi velocity $V_{F}=10^{15}nm/s$. We find rates of  $Im[\Pi^{rpa}(\boldsymbol{q},\hbar \omega)]e^{-2qz}$ ranging from ($-10^{5}$ $eV^{-1}nm^{-2}$ to $0 \,eV^{-1}nm^{-2}$) for z=5nm. We also show In Fig.\ref{fig:1}-(a) the electron-hole continuum or single particle excitation regions in $(q,E_{X})$ space, which determines the absorption (Landau damping) of the external field at given frequency and wave vector. The single particle excitation continuum is defined by the non-zero value of the imaginary part of the polarizability function, $Im[\Pi^{rpa}(\boldsymbol{q},\hbar \omega)] \neq 0$, and for the graphene both intraband and interband transition of the single particle excitation are possible, and the boundaries are as showing in the inset of Fig. \ref{fig:1}-(a) . If the collective mode lies inside the single particle excitation continuum we expect the mode to be damped. For graphene the plasmon lies inside the interband single particle excitation continuum decaying into electron-hole pairs. Only in the region 1A of fig.\ref{fig:1}-(a) that the plasmon is not damped, also the graphene plasmon does not enter into the intraband single particle excitation and it exists for all wave vectors. The highest absorption rate is observed in the single particle excitation interband regime (2B region), for the lowest excitation energy and for $q \rightarrow 0$ where the exponent $e^{-2qz}$ in this term decays rapidly with q, which guarantees that when $\hbar \omega$ is fixed the dominant contribution to this term comes from lowest possible q where the imaginary part of the polarization operator is still nonzero. We also mention that the optical absorption of the graphene layer decrease as a function of the increase of the interlayer distance . Now to investigate the dependence of quenching rate $\Gamma_{q}$ on the excitation energy we show at figure \ref{fig:1}-(b) the quenching rate for the adopted excitation energy of the $WSe_{2}$ QD like with a fixed interlayer distance (z=5nm) multiplied by the Dirac function $\delta(\hbar \omega-E_{X})$ where we obtain a photo-luminescence replica with an intensity corresponding to the quenching rate $(8.54 ns^{-1})$ for that given distance and energy. In fig.\ref{fig:1}-(c) we investigate the variation of the quenching rate in function of the wave vector q  where it represents a cut of the quenched polarisibility of figure \ref{fig:1}-(a) for a fixed excitation energy $E_{X}=1.654eV$.  We also find that the FRET quenching rate has a maximum at q=0.6 nm and it shape is governed by the $Im[\Pi^{rpa}(\boldsymbol{q},E_{X})]$ function that is relative for two point charges located at the innermost portion  of the graphene sheet and for very small wavevectors, $Im[\Pi^{rpa}(\boldsymbol{q},E_{X})] \approx 0$ (there is no screening inside the sheet) since very long wavelength perturbations correspond to charges being so far apart that they only experience the screening of the surrounding environment. For increasing wavevector q, the charges get closer, so more of the field lines connecting them are confined to the interior of the graphene sheet. we conclude that $Im[\Pi^{rpa}(\boldsymbol{q},E_{X})]$ increases linearly with q until it reaches near a maximum, and subsequently decreases with increasing q after the maximum due to the intrinsic inability of the medium to screen very short-wavelength excitations.

\subsection{The environment effect on the SPE's quenched radiative lifetime }

 \begin{figure}[ht]%
 \begin{center}
 \includegraphics[scale=0.35]{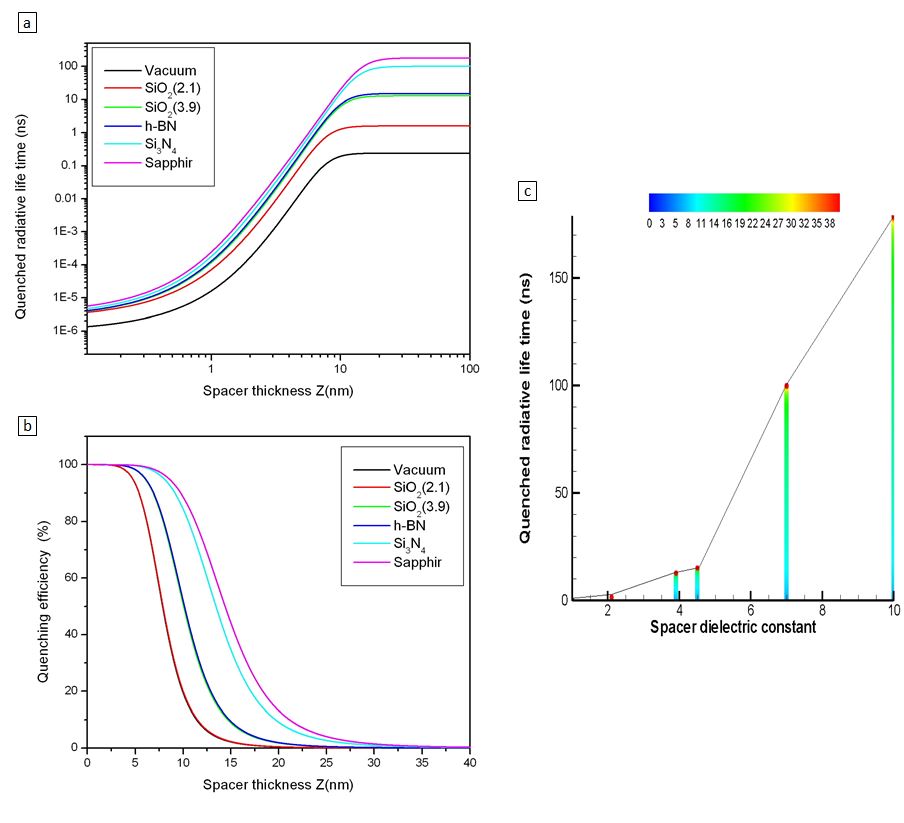}
 \caption{(a)and(b)Numerical evaluation of quenched radiative lifetime of the TMD QD like system as a function of the interlayer distance (spacer thickness) and the spacer dielectric constant epsilon,(c) the FRET efficiency in function of the the interlayer distance and spacer dielectric constant. } \label{fig:2}
 \end{center}
 \end{figure}
 
We investigate the effect of the spacer dielectric constant on the SPE's quenched radiative lifetime   in a way to gain control over it. To get a first impression of this effect we exploit Eq.\ref{eq:8} where the dielectric constant of the spacer interferes in the quenched radiative lifetime through the energy given by the QD-like $WSe_{2}$, its relative  oscillator strength and finally through the screened excitations of the graphene translated by the RPA approximation. In figure \ref{fig:2} we evaluate all possible spacers materials, this will give us a control over the desired radiative lifetime where the dielectric constant is considered as a gate to tone the apparent radiative lifetime of the tmd material. This quenched radiative lifetime of the SPE is shown in figure \ref{fig:2}-(a) in function of the spacer thickness for different dielectric constant while in figure \ref{fig:2}-(c) we try to extrapolate the general behaviour of this SPE's quenched radiative lifetime in function of the dielectric constant taking into  account the effect of $ \varepsilon_{spacer} $ on the RPA approximation and  for a spacer thickness z ranging from 0 nm to 100 nm. We find that This Förster induced radiative lifetime exhibits a smooth decay in function of the spacer dielectric constant  ranging from ($\tau^{sapphir}_{max}=$ 179 ns ; $\tau^{sapphir}_{min}=$ 3.26 fs) relative to the lowest excitation energy which correspond to the highest spacer dielectric constant that of the sapphir spacer to ($\tau^{vacuum}_{max}=$ 0.24 ns ; $\tau^{vaccum}_{min}=0.968$ fs) for the highest accessible excitation energy corresponding to the lowest spacer constant and that is of the vacuum. We show in fig \ref{fig:2}-b, that changing the spacer material can pushes the limit of the quenching distance  by 12 nm and we can mention that the lowest efficiency by distance curve is observed for the $vaccum$ and The biggest one is for the $sapphir$. We attribute this change to the direct effect of spacer dielectric constant on the SPE's energy and its corresponding radiative lifetime and also on the screening in the graphene sheet.
 
 \begin{table}[ht]%
\caption{\label{tab:table1}
The used spacer dielectric constant for determining the ML $WSe_{2}$ relative SPE energy $E_{X}$, radiative lifetime $\tau_{max}$
} and its correspondent quenched radiative lifetime $\tau$ and quenching effeciency $\varphi_{eff}$ for z=6 nm and z=10 nm.
\begin{ruledtabular}
\begin{tabular}{c|cccccc}
Substrat&Sapphir&$Si_{3}N_{4}$&h-BN&$SiO_{2}$&$SiO_{2}$&vacuum\\
\hline
$\epsilon_{spacer}$&10&7&4.5&3.9&2.1&1\\
$E_{X} (eV)$&2.107&2.064&1.967&1.922&1.65&1.317\\
$\tau_{max}(ns)$&179&100&15&12.8&1.6&0.24\\
$\tau(z=6nm)(ns)$&1.37 & 1.05 & 0.72 & 0.63& 0.27 &0.04\\
$\varphi_{eff}(z=10nm) \%$& 88.5 & 84.3 & 50.6 & 50.6 & 19 & 19 \\
\end{tabular}
\end{ruledtabular}
\end{table}

\subsection{SPE's quenched radiative lifetime in the presence of different TMD's}

\begin{figure}[ht]%
 \begin{center}
 \includegraphics[scale=0.25]{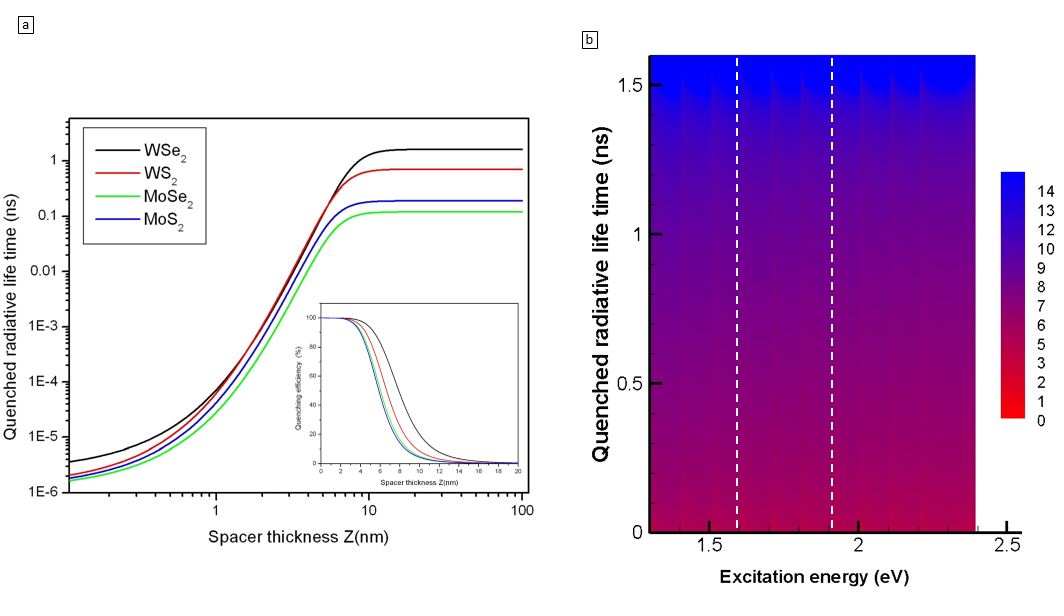}
 \caption{(a) numerical calculation of the the FRET radiative lifetime in function of the interlayer distance for the different TMD materials on an Sio2 substrate with an inset of their relative quenching efficiency , (b)the general behaviour of the quenched radiative lifetime in function of the excitation energy and the interlayer distance.} \label{fig:3}
 \end{center}
 \end{figure}
 
 \begin{table}[ht]%

\caption{\label{tab:table2}
Different ML TMD materials diposted on a $SiO_{2}$ substrate and their relative electron and hole effective masses  $m_{e(h)}$, $E_{g}$ gap energy, correspondent SPE energy $E_{X}$ and radiative lifetime $\tau_{max}$ we also provide the quenched radiative lifetime $\tau$ and the quenching efficiency $\varphi_{eff}$ for a z=5nm spacer thickness
}
\begin{ruledtabular}
\begin{tabular}{c|cccc}
$TMD/SiO_{2}$&$WSe_{2}$&$WS_{2}$&$MoSe_{2}$&$MoS_{2}$\\
\hline
$m_{e} (m_{0})$ \cite{F51,F52,F53,F54}&0.29&0.31&0.5&0.45\\
$m_{h} (m_{0})$ \cite{F51,F52,F53,F54}&0.36&0.42&0.6&0.54\\
$E_{g} (eV)$ \cite{F55} &2.08&2.43&2.18&2.48\\
$E_{X} (eV)$&1.65&1.99&1.544&1.77\\
$\tau_{max}(ns)$& 1.6 & 0.7 &0.12 & 0.19\\
$\tau(z=6nm)(ps)$& 275 & 249 & 59 &  101\\
$\varphi_{eff}$(z=5nm)($\%$) & 83 &83 & 74 & 71 \\
\end{tabular}
\end{ruledtabular}

\end{table}

We are going to shed the light on the effect of the graphene sheet on other TMD's SPE's. Firstly we should mention that the big difference between these  materials in our frame work consists mainly in the electron and hole effective masses which are given in the table \ref{tab:table2}, secondly we show the difference in the SPE's radiative lifetime of TMD sheets (for a $SiO_{2}$ substrate) when the system is isolated from the graphene sheet, $\tau^{WSe_{2}}_{max}-\tau^{MoSe_{2}}_{max}=$ 1.41 ns  where the biggest radiative lifetime is attribuet to ML $WSe_{2}$ $\tau^{WSe_{2}}_{max}=$ 1.6 ns or the lowest one is attribute for $MoSe_{2}$ $\tau^{MoSe_{2}}_{max}=$ 0.15 ps. In the presence of graphene and in close proximity the difference of the radiative lifetime becomes in order of $\tau^{WSe_{2}}_{min}-\tau^{MoSe_{2}}_{min}=1.28$ fs where the lowest is attribute to the $MoSe_{2}$ with $\tau^{MoSe_{2}}_{min}=1.12$ fs and the highest is for the $WSe_{2}$ with $\tau^{WSe_{2}}_{min}=2.4$ fs . Changing the TMD material has proved ,in the inset of fig.\ref{fig:3}-a, that it pushes the limit of the quenching distance  by 4-6 nm and we can mention that the lowest efficiency by distance curve is observed for ML $MoSe_{2}$  while The biggest one is for the $WSe_{2}$ ML.  the dominant variable here affecting this radiative lifetime is the SPE's energy, therefore we only take it in consideration and we extrapolate and plot the quenched radiative lifetime fig \ref{fig:3}-b. in function of the SPE's energy and the interlayer distance where we suppose that ($1/\Gamma_{TMD}= 1.6$ ns). We can clearly see that the quenching efficiency has a unique dependence for the excitation energy where it's responsible for the the curve shape and we find that for low SPE's energy (1.3 eV range) and for high ones( over 1.9eV), we got a brutal change of the quenched radiative lifetime from low $\tau$ values to big $\tau$ values in function of the distance but for SPE's energy between 1.6 eV and 1.9 eV we have a smooth change of $\tau$ in function of the distance.

\section{Conclusion}

In conclusion, we have presented a simple analytic model describing the single photon emission in ML TMD and the corresponding Förster induced radiative lifetime in the hybrid heterostructure consisting of graphene, dielectric spacer and  ML TMD. We predict that the Förster quenching rates leads to a low radiative lifetime of this Single photon emission while we gain control over this quenching through two major gates, the dielectric spacer nature and the choice of the ML TMD material. Our work predict that the corresponding SPE’s quenched radiative lifetime will be in the picoseconds range for a given distance, this time scale is in good agreement with the recently measured exciton lifetime in these heterostructures [7-10]. 

\begin{acknowledgments}
The authors gratefully acknowledge fruitful and stimulating discussions with Haithem zahra.
\end{acknowledgments}

\appendix

\section{Localised exciton states }
To estimate the energies and the oscillator strength of the localized  exciton, we use 2D effective-mass approximation. It is convenient to work in the center of- mass frame and the physics of exciton is described by the following shr\"{o}dinguer equation:
\begin{align}
\bigg[E_{g}-\dfrac{\hbar^{2}\nabla^{2}_{\boldsymbol{\rho}}}{2\mu}+V_{ky}(\boldsymbol{\rho})-\dfrac{\hbar^{2}\nabla^{2}_{\boldsymbol{R}}}{2 M_{X}}+ V(\boldsymbol{R})\bigg] \Psi_{X}(\boldsymbol{\rho},\theta,\boldsymbol{R}) \notag \\ =E_{X} \Psi_{X}(\boldsymbol{\rho},\theta,\boldsymbol{R})
\end{align}
with $E_{g}$ is the gap energy and $V(R)$ is the disorder potential derived from structural imperfections given by Eq.1 in the main text, the electron-hole direct Coulomb interaction is treated here using the Rytova-Keldysh potential $V_{ky}(\rho)$ according to the widely accepted approach.$\mu=\frac{m_{e}m_{h}}{M_{X}}$,  is the reduced effective mass, $M_{X}$ is the exciton mass, $m_{e}$ ($m_{h}$) is the electron (hole) effective mass, given in terms of free electron effective mass units ($m_{0}$).

In our work , it is assumed that the perturbation introduced by disorder is not sufficient to produce a transition from the exciton 1s state to higher states of the relative electron-hole
motion. Hence, the exciton stays always in the 1s state and only its center of mass motion is affected by disorder. We make the following factorization ansatz for the localised exciton wave function solution of Eq A1:
\begin{align}
\Psi_{X}(\boldsymbol{\rho},\theta,\boldsymbol{R})=\phi_{\tilde{nl}}(\boldsymbol{\rho})  \times \psi^{CM}_{n_{x},n_{y}}(\boldsymbol{R})
\end{align}
where $\phi_{\tilde{nl}}(\boldsymbol{\rho})=\sum _{n, \vert l \vert < n} C(n,l) \varphi _{n,l}(\boldsymbol{\rho})$ are the eigenvalue solution of the relative hamiltonian. $\phi_{\tilde{nl}}(\boldsymbol{\rho})$ expanded in terms of 2D-hydrogenic states $\varphi _{n,l}(\boldsymbol{\rho})$. $ n = 1, 2, 3... $ is the principal quantum number, $l= 0, \pm 1, \pm 2, \pm 3,...\pm n-1$ is the angular momentum number and  $\psi^{CM}_{n_{x},n_{y}}(\boldsymbol{R})=\sum_{n_{x},n_{y}} D(n_{x},n_{y})\frac{1}{\sqrt{n_{x}!n_{y}!}}\frac{1}{\sqrt{2^{n_{x},n_{y}}}}\sqrt{\frac{1}{\pi R_{cm} ^{2}}}\\ \mathbf{H}_{n_{x}} (\frac{X}{R_{cm}}) 
\mathbf{H}_{n_{y}}(\frac{Y}{R_{cm}})e^{-\frac{(X^{2}+Y^{2})}{2R_{cm}^{2}}}$ is the center of mass wave function, obtained by the numerical diagonalization of the matrix resulting from the projection of the center of mass Hamiltonian on the basis of 2D harmonic oscillator. $R_{cm}=\sqrt{\frac{\hbar}{M_{X}\omega_{cm}}}$ is the center of mass  localization radius in the harmonic potential description and $\omega_{cm}$ is the frequency of a 2D harmonic oscillator.$\mathbf{H}_{n_{x(y)}} $ is the Hermite polynomials. In this notation, the integers $n_{x}$, $n_{y}$ are quantum numbers.\\
The solution of the Schrodinger equation for a localized exciton in a
solid is
\begin{align*}
\zeta_{X} (\boldsymbol{R},\boldsymbol{\rho},z_{e},z_{h})=\Psi_{X}(\boldsymbol{\rho},\theta,\boldsymbol{R}) u_{c,k}(\boldsymbol{r_{e}}) u_{v,k}^{*}(\boldsymbol{r_{h}})   
\end{align*}

$u_{c,k}(\boldsymbol{r_{e}})$ and $u_{v,k}^{*}(\boldsymbol{r_{h}})$ are the Bloch function taken at the points of high symmetry ($K_{\tau}, \tau=\pm1$) of the valance(v) and conduction (c) bands. In our case, we neglect the possible mixtures with the other excitonic states.
In order to calculate the optical matrix element obtained by considering the interaction with the first order electromagnetic field, it is convenient to introduce the Fourier transforms of the envelope-function $\phi_{n,l}(\rho,\theta)$ and $\psi_{n_{x},n_{y}}^{CM}(\boldsymbol{R})$ defined in the plane of the two-dimensional  crystal:

\begin{align}
\psi_{n_{x},n_{y}}^{CM}(\boldsymbol{R})=\dfrac{1}{(2\pi)^{2}}\int \int e^{i\boldsymbol{K}.\boldsymbol{R}} \psi_{n_{x},n_{y}}^{CM}(\boldsymbol{K})d^{2}\boldsymbol{K}
\end{align}

\begin{align}
\phi_{n,l}(\boldsymbol{\rho})=\dfrac{1}{(2\pi)^{2}}\int \int e^{i\boldsymbol{k}.\boldsymbol{\rho}} \phi_{n,l}(\boldsymbol{k}) d^{2}\boldsymbol{k}
\end{align}
we thus obtain the alternative expression:
 \begin{align}
\zeta_{X} (\boldsymbol{R},\boldsymbol{\rho},z_{e},z_{h}) =\dfrac{1}{(2\pi)^{4}}\int\int \int \int d^{2}\boldsymbol{K} d^{2}\boldsymbol{k} \notag \\
e^{i \big(\boldsymbol{K}.\boldsymbol{R}+\boldsymbol{k}.\boldsymbol{\rho}\big)} \psi_{n_{x},n_{y}}^{CM}(\boldsymbol{K}) \phi_{n,l}(\boldsymbol{k})  u_{c,k}(\boldsymbol{r_{e}}) u_{v,k}^{*}(\boldsymbol{r_{h}})
\end{align}

whith $\boldsymbol{K}= \boldsymbol{k_{e}}+\boldsymbol{k_{h}}$ , $\boldsymbol{k}= \dfrac{m_{h}\boldsymbol{k_{e}}-m_{e}\boldsymbol{k_{h}}}{M_{X}}$. Use  the following change of variable ($\boldsymbol{K}, \boldsymbol{k}) \leadsto (\boldsymbol{k_{e}}, \boldsymbol{k_{h}})$ we can write:

 \begin{align}
\zeta_{X} (\boldsymbol{R},\boldsymbol{\rho},z_{e},z_{h})=\dfrac{1}{(2\pi)^{4}}\int\int \int \int d^{2}\boldsymbol{k_{e}} d^{2}\boldsymbol{k_{h}} \notag \\ e^{i \big(\boldsymbol{k_{e}}.\boldsymbol{\rho_{e}}+\boldsymbol{k_{h}}.\boldsymbol{\rho_{h}}\big)}  \psi_{n_{x},n_{y}}^{CM}(\boldsymbol{K})\phi_{n,l}(\boldsymbol{k})  u_{c,k}(\boldsymbol{r_{e}}) u_{v,k}^{*}(\boldsymbol{r_{h}})
\end{align}
we can then pass to discrete summations on the allowed states of the first Brillouin zone, because $\vert\boldsymbol{K} \vert \sim \dfrac{1}{L} \ll \vert\boldsymbol{b_{1}}\vert, \vert\boldsymbol{b_{2}}\vert $, $\vert\boldsymbol{k}\vert \sim \dfrac{1}{a_{B}} \ll \vert\boldsymbol{b_{1}}\vert, \vert\boldsymbol{b_{2}}\vert$
where $\boldsymbol{b_{1}}$, $\boldsymbol{b_{2}}$  are the basis vectors of the reciprocal 2D lattice. 
$\boldsymbol{b_{1}}$ and $\boldsymbol{b_{2}}$ are the basis vectors of the reciprocal 2D lattice.
\begin{align}
\zeta_{X} (\boldsymbol{R},\boldsymbol{\rho},z_{e},z_{h})=\dfrac{1}{S^{2}} \sum_{\boldsymbol{k_{e}}, \boldsymbol{k_{h}} \in ZB} e^{i \big(\boldsymbol{k_{e}}.\boldsymbol{\rho_{e}}+\boldsymbol{k_{h}}.\boldsymbol{\rho_{h}}\big)} \psi_{n_{x},n_{y}}^{CM}(\boldsymbol{K}) \notag \\ \phi_{n,l}(\boldsymbol{k})  u_{c,k}(\boldsymbol{r_{e}}) u_{v,k}^{*}(\boldsymbol{r_{h}})
\end{align}
The Bloch functions $u_{c,k}(\boldsymbol{r_{e}})$, $u_{v,k}^{*}(\boldsymbol{r_{h}})=u_{h,-k}(\boldsymbol{r_{h}})$ vary  slowly when $\boldsymbol{k_{e}}$($\boldsymbol{k_{h}}$) varies around the K point. we can write then :
\begin{align}
\zeta_{X} (\boldsymbol{R},\boldsymbol{\rho},z_{e},z_{h})=\dfrac{1}{S} \sum_{\boldsymbol{k_{e}}, \boldsymbol{k_{h}} \in ZB} \psi_{n_{x},n_{y}}^{CM}(\boldsymbol{K})\phi_{n,l}(\boldsymbol{k})  \notag \\ \Upsilon_{c,k_{e}}(\boldsymbol{r_{e}}) \Upsilon_{v,k_{h}}^{*}(\boldsymbol{r_{h}})
\end{align}
where :
\begin{eqnarray}
\Upsilon_{c,k_{e}}(\boldsymbol{r_{e}})=\dfrac{1}{\sqrt{S}} e^{i \boldsymbol{k_{e}}.\boldsymbol{\rho_{e}}} u_{c,\boldsymbol{k_{e}}}(\boldsymbol{r_{e}})
\end{eqnarray}

\begin{eqnarray}
\Upsilon_{v,k_{h}}(\boldsymbol{r_{h}})=\dfrac{1}{\sqrt{S}} e^{i \boldsymbol{k_{v}}.\boldsymbol{\rho_{h}}} u_{v,\boldsymbol{k_{h}}}(\boldsymbol{r_{h}})
\end{eqnarray}
here $\boldsymbol{k_{h}}=-\boldsymbol{k_{v}}$
In Fock representation, the electromagnetic field is written:
\begin{align}
\hat{\boldsymbol{A}}_{\boldsymbol{q}}^{\lambda}(\boldsymbol{r},t)=\bigg(\dfrac{2\pi\hbar c}{n_{0}q V}\bigg) \bigg[\boldsymbol{\epsilon}_{\boldsymbol{q}}^{\lambda} e^{i\big(\boldsymbol{q}.\boldsymbol{r}-\omega_{\boldsymbol{q}} t\big)} a_{\boldsymbol{q},\lambda} + \boldsymbol{\epsilon}_{\boldsymbol{q}}^{\lambda} e^{-i\big(\boldsymbol{q}.\boldsymbol{r}-\omega_{\boldsymbol{q}} t\big)} a_{\boldsymbol{q},\lambda}^{+}\bigg]
\end{align}
$\boldsymbol{\epsilon}_{\boldsymbol{q}}^{\lambda}$  is a unit vector characterizing the optical mode polarization $\lambda$. The operator $a_{\boldsymbol{q},\lambda}^{+}$ ($a_{\boldsymbol{q},\lambda}$) creates (annihilates)
a photon with wave vector $\boldsymbol{q}$ and polarization $\lambda$. The optical angular frequency is defined by $\omega_{\boldsymbol{q}}=\dfrac{c}{n_{0}} \vert\boldsymbol{q}\vert$, here  $n_{0}$ is the effective optical refraction index of the crystal environment, $c$ is the light velocity and $V$ is the normalization volume. We treat the modification of the 2D system by the fields using the first order time-dependent perturbation  theory , by ignoring the terms in $\hat{\boldsymbol{A}}^{2}$, i.e. for sufficiently weak electromagnetic fields. So, in Coulomb gauge ($div \hat{\boldsymbol{A}}=0$):
\begin{align}
\hat{H}_{opt}^{\lambda}=\dfrac{e \hat{\boldsymbol{A}}(\boldsymbol{r})}{m_{0}c}\sum_{i}\hat{\boldsymbol{p}}_{i}
\end{align}
$\boldsymbol{p}$ is the electron momentum operator, the sum is expected at all electrons in the system.
\begin{align}
\hat{\boldsymbol{A}}_{\boldsymbol{q}}^{\lambda}(\boldsymbol{r},t)= \hat{\boldsymbol{A}}_{\boldsymbol{q}}^{\lambda}(\boldsymbol{r}) e^{-i\omega_{\boldsymbol{q}} t}+ \hat{\boldsymbol{A}}_{\boldsymbol{q}}^{+\lambda}(\boldsymbol{r}) e^{i\omega_{\boldsymbol{q}} t} 
\end{align}
The optical matrix element characterizing the transition from the 2D crystal ground state to the exciton state can be writen as follow :
\begin{align}
M_{\boldsymbol{q},\lambda}= \bigg(\dfrac{2\pi\hbar c}{n_{0}q V}\bigg)^{\dfrac{1}{2}} \langle \Psi_{0} \vert e^{i\boldsymbol{q}.\boldsymbol{r}} \boldsymbol{\epsilon}_{\boldsymbol{q}}^{\lambda} \sum_{i}\hat{\boldsymbol{p}}_{i}\vert \zeta_{X} \rangle
\end{align}

\begin{align}
M_{\boldsymbol{q},\lambda}=  \bigg(\dfrac{2\pi\hbar c}{n_{0}q V}\bigg)^{\dfrac{1}{2}} \boldsymbol{\epsilon}_{\boldsymbol{q}}^{\lambda} \langle \Psi_{0} \vert\sum_{i}\hat{\boldsymbol{p}}_{i}\vert \zeta_{X} \rangle
\end{align}

In gauge Coulomb  $[e^{i\boldsymbol{q}.\boldsymbol{r}},\hat{\boldsymbol{p}}_{i}]=0$. We define the oscillator strength of a transition by the dimensionless expression:

\begin{align}
f_{\boldsymbol{\epsilon}_{\boldsymbol{q}}^{\lambda}}= \dfrac{2}{m_{0}\hbar\omega_{0}} \vert \bar{M}_{\boldsymbol{q},\lambda} \vert
\end{align}
where $\omega_{0}=\vert\omega_{i}-\omega_{f}\vert$ here , $\omega_{0}=\omega_{X}$ and 
\begin{align}
\bar{M}_{\boldsymbol{q},\lambda}=\boldsymbol{\epsilon}_{\boldsymbol{q}}^{\lambda} \langle \Psi_{0} \vert\sum_{i}\hat{\boldsymbol{p}}_{i}\vert \zeta_{X} \rangle
\end{align}
\begin{align}
\bar{M}_{\boldsymbol{q},\lambda} &=  \dfrac{1}{s}\sum_{\boldsymbol{k_{e}},\boldsymbol{k_{h}}} \psi_{n_{x},n_{y}}^{CM}(\boldsymbol{K})\phi_{n,l}(\boldsymbol{k}) \notag \\ & \langle \Upsilon_{c,k_{e}} \vert \langle \Upsilon_{v,-k_{h}}^{*}  \vert \boldsymbol{\epsilon}_{\boldsymbol{q}}^{\lambda}  .\sum_{i}\hat{\boldsymbol{p}}_{i}\vert \Psi_{0} \rangle
 = \dfrac{1}{S} \sum_{\boldsymbol{k_{e}},\boldsymbol{k_{h}}} \notag \\ & \psi_{n_{x},n_{y}}^{CM}(\boldsymbol{K})\phi_{n,l}(\boldsymbol{k})  \langle \Upsilon_{c,k_{e}} \vert   e^{i\boldsymbol{q}.\boldsymbol{r}} \boldsymbol{\epsilon}_{\boldsymbol{q}}^{\lambda} \hat{\boldsymbol{p}}\vert \Upsilon_{v,k_{v}} \rangle
\end{align}
since $\boldsymbol{k_{e}}$, $\boldsymbol{k_{v}}$ $\in$ Brillouin zone , and $\vert\boldsymbol{q} \vert\ll \vert \boldsymbol{b_{1}}\vert, \vert \boldsymbol{b_{2}}\vert$

$$\langle \Upsilon_{c,k_{e}} \vert   e^{i\boldsymbol{q}.\boldsymbol{r}} \boldsymbol{\epsilon}_{\boldsymbol{q}}^{\lambda} \hat{\boldsymbol{p}}\vert \Upsilon_{v,k_{v}} \rangle = \delta_{\boldsymbol{k_{e}}, \boldsymbol{k_{v}}+ \boldsymbol{q}} \langle u_{c,k_{\tau}} \vert  \boldsymbol{\epsilon}_{\boldsymbol{q}}^{\lambda} . \hat{\boldsymbol{p}}\vert u_{v,k_{\tau}} \rangle$$
hence
\begin{align}
\bar{M}_{\boldsymbol{q},\lambda} = \dfrac{1}{S} \sum_{\boldsymbol{k_{e}},\boldsymbol{k_{h}}} \psi_{n_{x},n_{y}}^{CM}(\boldsymbol{K})\phi_{n,l}(\boldsymbol{k}) \langle u_{c,k_{\tau}} \vert  \boldsymbol{\epsilon}_{\boldsymbol{q}}^{\lambda} . \hat{\boldsymbol{p}}\vert u_{v,k_{\tau}} \rangle \delta_{\boldsymbol{k_{e}}, \boldsymbol{k_{v}}+ \boldsymbol{q}} 
\end{align}
$\boldsymbol{K}=\boldsymbol{k_{e}}-\boldsymbol{k_{v}}=\boldsymbol{q}$ and $\boldsymbol{k}=\boldsymbol{k_{e}}-\dfrac{m_{e}}{M_{X}} \boldsymbol{q}$
or $\vert \boldsymbol{q} \vert \ll \dfrac{\pi}{L}$, on the scale of $\vert \boldsymbol{q} \vert$ $\psi_{n_{x},n_{y}}^{CM}(\boldsymbol{q})$ is slowly varying therefore $\psi_{n_{x},n_{y}}^{CM}(\boldsymbol{q})= \psi_{n_{x},n_{y}}^{CM}(0)$  
\begin{align}
\bar{M}_{\boldsymbol{q},\lambda} = \dfrac{1}{S}  \psi_{n_{x},n_{y}}^{CM}(\boldsymbol{K}=0)\sum_{\boldsymbol{k_{e}}} \phi_{n,l}(\boldsymbol{k_{e}})  \boldsymbol{\epsilon}_{\boldsymbol{q}}^{\lambda} \langle u_{c,k_{\tau}} \vert \hat{\boldsymbol{p}}\vert u_{v,k_{\tau}} \rangle 
\end{align}

or $$\sum_{\boldsymbol{k_{e}}} \phi_{n,l}(\boldsymbol{k_{e}})= \dfrac{S}{(2\pi)^{2}}\int\int d^{2}\boldsymbol{k_{e}} \phi_{n,l}(\boldsymbol{k_{e}}) = S \phi_{n,l}(\rho=0)$$
and 
$$ \psi_{n_{x},n_{y}}^{CM}(\boldsymbol{K}=0)= \int\int \psi_{n_{x},n_{y}}^{CM}(\boldsymbol{R}) d^{2}\boldsymbol{R}$$
finally 
\begin{align}
\bar{M}_{\boldsymbol{q},\lambda}= \bigg(\int\int \psi_{n_{x},n_{y}}^{CM}(\boldsymbol{R}) d^{2}\boldsymbol{R} \bigg) \phi_{n,l}(\rho=0)  \boldsymbol{\epsilon}_{\boldsymbol{q}}^{\lambda} \langle u_{c,k_{\tau}} \vert \hat{\boldsymbol{p}}\vert u_{v,k_{\tau}} \rangle
\end{align}
then we can rewriten the oscillator strength as follow :

\begin{align}
f_{\boldsymbol{\epsilon}_{\boldsymbol{q}}^{\lambda}}= \dfrac{2}{m_{0}\hbar \omega_{X}}  \bigg \vert \bigg(\int\int \psi_{n_{x},n_{y}}^{CM}(\boldsymbol{R}) d^{2}\boldsymbol{R}\bigg) \bigg\vert^{2}  \bigg\vert \phi_{n,l}(\rho=0) \bigg\vert^{2} \notag \\ \bigg \vert  \langle u_{c,k_{\tau}} \vert \boldsymbol{\epsilon}_{\boldsymbol{q}}^{\lambda}.\hat{\boldsymbol{p}}\vert u_{v,k_{\tau}} \rangle \bigg \vert^{2}
\end{align}
The oscillator strength can also be expressed in terms of the dipole matrix element as :
\begin{align}
f_{\boldsymbol{\epsilon}_{\boldsymbol{q}}^{\lambda}}= \dfrac{2 m_{0} \omega_{j}^{X}}{\hbar} \bigg\vert \bigg\langle \Psi_{0} \bigg \vert \boldsymbol{\epsilon}_{\boldsymbol{q}}^{\lambda} \sum_{i}\hat{\boldsymbol{r}}_{i} \bigg \vert \Psi_{X} \bigg\rangle \bigg\vert^{2}
\end{align}
For circularly polarized light $\sigma_{\pm}$ propagating along the normal to the sample ($p_{\pm}= \dfrac{p_{x}\pm i p_{y}}{2}$), so that only optical modes with in-plane polarization components couple to these excitons 
\begin{align}
\bigg \vert d_{\parallel} \bigg\vert^{2} &=e^{2} \bigg\vert \bigg\langle \Psi_{0}  \bigg \vert \boldsymbol{\epsilon}_{\boldsymbol{q}}^{\lambda} \sum_{i}\hat{\boldsymbol{r}}_{i} \bigg \vert \Psi_{X} \bigg\rangle \bigg\vert^{2}= \notag \\&  \dfrac{e^{2}}{ (m_{0} \omega_{X})^{2}} \bigg \vert \bigg(\int\int \psi_{n_{x},n_{y}}^{CM}(\boldsymbol{R}) d^{2}\boldsymbol{R} \bigg) \bigg\vert^{2}  \bigg\vert \phi_{n,l}(\rho=0) \bigg\vert^{2}  \notag \\& \bigg \vert  \langle u_{c,k_{\tau}} \vert \hat{\boldsymbol{p_{\pm}}}\vert u_{v,k_{\tau}} \rangle \bigg \vert^{2}
\end{align}

\section{ 2D-coulomb Fourier transform}
\subsection{The coulomb potential due to a point charge}

First, we find the 2D Fourier Transform of Coulomb potential, created in-plane, by a unit charge positioned at a distance z out of plane. Specific coordinates of the charge are $(x_{0}; y_{0}; z_{0})$ and the plane is defined by z = 0. The real-space Coulomb potential of the unit charge is

\begin{equation}
V(\textbf{r})=\dfrac{Q}{\varepsilon_{spacer}\sqrt{z_{0}^{2}+(x-x_{0})^{2}+(y-y_{0})^{2}}},
\end{equation}

where the in-plane vector $\boldsymbol{r}$ =(x; y; 0). Its 2D Fourier transform is

\begin{equation}
V(\textbf{q})=\frac{Q}{\varepsilon_{spacer}}\int d\boldsymbol{r} \dfrac{e^{-i(\textbf{q.r})}}{\sqrt{z_{0}^{2}+(x-x_{0})^{2}+(y-y_{0})^{2}}}.
\end{equation}

Substituting $ x - x0\rightarrow$ x and $ y - y0 \rightarrow$ y produces

\begin{equation}
V(\textbf{q})=\frac{Q e^{-i(\textbf{q}.\textbf{r}_{0})}}{\varepsilon_{spacer}}\int d\boldsymbol{r} \dfrac{e^{-i(\textbf{q.r})}}{\sqrt{z_{0}^{2}+x^{2}+y^{2}}},
\end{equation}

where $\boldsymbol{r_{0}}\, =\, (x_{0}; y_{0};0)$. Rewriting the integral equivalently in polar coordinates produces

\begin{equation}
V(\textbf{q})=\frac{Q \, e^{-i(\textbf{q}.\textbf{r}_{0})}}{\varepsilon_{spacer}}\int \boldsymbol{r} \,d\boldsymbol{r} \dfrac{e^{-i(\textbf{q.r})}}{\sqrt{z_{0}^{2}+\textbf{r}^{2}}} \int d\theta e^{-iqr \, cos\,\theta}.
\end{equation}

The angular integral is evaluated (in Mathematica, or using integral tables by, for example, Gradshteyn and Ryzhik) to produce

\begin{equation}
\int_{0}^{2\pi}  e^{-iqr \, cos\,\theta} d\theta \,= \, 2 \pi J_{0}(\boldsymbol{q.r})
\end{equation}

where $J_{0}(x)$ is the Bessel function of the first kind. This results in

\begin{equation}
V(\textbf{q})=\frac{ 2 \pi \, Q}{\varepsilon_{spacer}} e^{-i(\textbf{q}.\textbf{r}_{0})}\int \boldsymbol{r} \, d\boldsymbol{r} \dfrac{e^{-i(\textbf{q.r})}}{\sqrt{z_{0}^{2}+\textbf{r}^{2}}}J_{0}(\boldsymbol{q.r}).
\end{equation}

This integral is also evaluated in Mathematica or using Gradshteyn $\&$ Ryzhik to produce

\begin{equation}
V(\textbf{q})=\frac{ 2 \pi \, Q}{\varepsilon_{spacer} \, q} e^{-i(\textbf{q}.\textbf{r}_{0})}e^{-qz_{0}}
\end{equation}

\subsection{The coulomb potential due to a DIPOLE}
The simplest way to obtain an in-plane potential of a dipole, located out of plane, is to use the result for point charge Eq. (7), and treat dipole as a collection of point charges. We substitute $ z_{0} \rightarrow z + \delta z_{i}$, and similarly for $x_{0}$ and $y_{0}$. Then, the total potential of collection of charge $Q_{i}$ with coordinates (x +$\delta x_{i}$; y + $\delta y_{i}$; z +  $\delta z_{i}$) is

\begin{equation}
V_{dip}(\textbf{q})=\frac{2\pi}{\varepsilon_{spacer} q} \sum_{i} Q_{i} e^{-q(z+\delta z_{i})} e^{-iq_{x}(x+\delta x_{i})} e^{-iq_{y}(y+\delta y_{i})}
\end{equation}

Assuming $\delta x_{i}$, $\delta y_{i}$ and $\delta z_{i}$ small and performing the Taylor expansion of exponents yields

\begin{align}
V_{dip}(\textbf{q})=\frac{2\pi} {\varepsilon_{spacer} q} \bigg[ \sum_{i} Q_{i} e^{-qz} e^{-iq_{x}x} e^{-iq_{y}y} \notag \\ + \sum_{i} Q_{i} (-q\delta z_{i}-iq_{x} \delta x_{i}-iq_{y}\delta y_{i} ) e^{-qz} e^{-iqr} \bigg]
\end{align}

The first r.h.s. term disappears once we assume that $\sum_{i} Q_{i}=0$ as is always the case for dipoles. The expression then becomes

\begin{equation}
V_{dip}(\textbf{q})=\frac{2e\pi}{\varepsilon_{spacer} \,q} (-qd_{z}-iq_{x}  d_{x}-iq_{y} d_{y} ) e^{-qz} e^{-i\boldsymbol{q.r}} 
\end{equation}

where $d\alpha$, $\alpha$ = x; y; z are components of the dipole vector. It straightforwardly transformed into

\begin{equation}
V_{dip}(\textbf{q})=\frac{2e\pi i}{\varepsilon_{spacer} } (-id_{z}-\frac{q_{x}  d_{x}}{q}-\frac{q_{y} d_{y}}{q} ) e^{-qz} e^{-i\boldsymbol{q.r}} 
\end{equation} 

Further, we re-write the scalar product as $q_{x}  d_{x} + q_{y} d_{y} = qd_{\parallel}\, cos \, \theta$, where $d_{\parallel} = (d_{x}; d_{y}; 0)$ and $\theta$ is the angle between q and $d_{\parallel}$. The result is then

 \begin{equation}
V_{dip}(\textbf{q})=\frac{2e\pi i}{\varepsilon_{spacer} } (-id_{z}-d_{\parallel}  cos \, \theta ) e^{-qz} e^{-i\textbf{q.r}} 
\end{equation}

This expression is very similar to Eq. (A7) in 2011 paper, except for some sign differences in the brackets, related to certain conventions of how the dipole is positioned with respect to the plane. The specific signs become irrelevant when calculating the rate of energy transfer, since the energy transfer will only depend on $|V(\textbf{q})|^{2}$. Another way to think about it is that the sign of specific projects of a transition dipole must not matter when obtaining results for experimentally observable rate since, semi-classically, the transition dipole is the amplitude of an oscillating dipole.


\begin{thebibliography}{[1]}

\bibitem{F1}
C. Janisch, Y. Wang, D. Ma, N. Mehta, A. L. Elías, N. Perea-Lopez, M. Terrones, V. Crespi, and Z. Liu, Sci. Rep.\textbf{4}, 5530 (2014).
\bibitem{F2}
 P. Tonndorf, R. Schmidt, R. Schneider, J. Kern, M. Buscema, G. A. Steele, A. Castellanos-Gomez, H. S. J. van der Zant, S. M. de Vasconcellos, and R. Bratschitsch, Optica \textbf{2}, 347 (2015).
\bibitem{F3}
B. Zhu, X. Chen, and X. Cui, Sci. Rep. \textbf{5}, 9218 (2015).
\bibitem{F4}
J. Huang, T. B. Hoang, andM. H.Mikkelsen, Sci. Rep. \textbf{6}, 22414 (2016).
\bibitem{F5}
A. T. Hanbicki, G. Kioseoglou, M. Currie, C. Stephen Hellberg, K. M. McCreary, A. L. Friedman, and B. T. Jonker, Sci. Rep. \textbf{6}, 18885 (2016).
\bibitem{F6}
A. R. Klots, A. K. M. Newaz , Bin Wang, D. Prasai , H. Krzyzanowska , Junhao Lin, D. Caudel, N. J. Ghimire, J. Yan , B. L. Ivanov , K. A. Velizhanin , A. Burger , D. G. Mandrus, N. H. Tolk , S. T. Pantelides and  K. I. Bolotin. Sci. Rep. \textbf{4}, 6608 (2014).
\bibitem{F7}7
A. Chernikov, T. C. Berkelbach, H. M. Hill, A. Rigosi, Y. Li, O. B. Aslan, D. R. Reichman, M. S. Hybertsen, and T. F. Heinz Phys. Rev. Lett. \textbf{113}, 076802 (2014).
\bibitem{F8}
A.T Hanbicki, M. Currie, G.Kioseogloub, A.L. Friedmana, B.T.Jonker, Solid State Commun. \textbf{203}, 16–20 (2015).
 \bibitem{F9}
C. Poellmann,  P. Steinleitner,  U. Leierseder,  P. Nagler, G. Plechinger,  M. Porer, R. Bratschitsch,  C. Schuller, T. Korn  and R. Huber,Nat. Mater. \textbf{14}, 889–894 (2015).
\bibitem{F10}
J. J. P. Thompson, S. Brem, H. Fang, J. Frey, S. P. Dash, W. Wieczorek, and E. Malic,  Phys. Rev. Materials \textbf{4}, 084006 (2020).
\bibitem{F11}
A.t Srivastava, M. Sidler, A. V. Allain , D. S. Lembke, A. Kis and A. Imamoglu, Nature Nanotechnology \textbf{10},  491–496 (2015).
\bibitem{F12}
P. Michler, A. Kiraz, C. Becher, W. Schoenfeld, P. Petro, L. Zhang, E. Hu, and A. Imamoglu, Science \textbf{290}, 2282 (2000). 
\bibitem{F13}
I. Aharonovich , S. Castelletto , D. A. Simpson , C.H. Su , A. D. Greentree , S. Prawer , Rep. Prog. Phys. \textbf{74}, 076501 (2011).
\bibitem{F14}
 P. Farrera, G. Heinze, B. Albrecht, M. Ho, M. Chavez, C. Teo, N. Sangouard, and H. De Riedmatten, Nat. Commun. \textbf{7}, 13556 (2016).
\bibitem{F15}
D. C. Burnham and D. L. Weinberg, Phys. Rev. Lett. \textbf{25}, 84 (1970).
\bibitem{F16}
T. T. Tran, K. Bray, M. J. Ford, M. Toth, and I. Aharonovich, Nat. Nano. \textbf{11}, 37 (2016).
\bibitem{F17}
N. Chejanovsky, M. Rezai, F. Paolucci, Y. Kim, T. Rendler, W. Rouabeh, F. Fvaro de Oliveira, P. Herlinger, A. Denisenko, S. Yang, I. Gerhardt, A. Finkler, J. H. Smet, and J. Wrachtrup, Nano Lett. \textbf{16}, 7037 (2016).
\bibitem{F18}
G. Grosso, H. Moon, B. Lienhard, S. Ali, D. K. Efetov, M. M. Furchi, P. J.Herrero, M. J. Ford, I. Aharonovich, and D. Englund, Nat. Commun. \textbf{8}, 705 (2017).
\bibitem{F19}
 H. J. Kimble, Nature \textbf{453}, 1023–1030 (2008).
\bibitem{F20}
A. Beveratos, R. Brouri, T. Gacoin, A. Villing, J.P. Poizat, and P. Grangier, Phys. Rev. Lett. \textbf{89}, 187901 (2002).
\bibitem{F21}
E. Knill, R. La, and G. J. Milburn ,Nature \textbf{409}, 46–52 (2001).
\bibitem{F22}
S. V. Polyakov and A. L. Migdall,  J. Mod. Opt. \textbf{56}, 1045–1052 (2009).
\bibitem{F23}
M. J. Stanley, C. Matthiesen, J. Hansom, C. Le Gall, C. H. H. Schulte, E. Clarke, and M. Atature,  Phys. Rev. B, \textbf{90}, 195305 (2014).
\bibitem{F24}
R. Stockill1, C. Le Gall, C. Matthiesen, L. Huthmacher, E. Clarke, M. Hugues and M. Atature, Nat. commun,  \textbf{7}, 1-7 (2016).
\bibitem{F25}
Geim, A. K.,  Novoselov, K. S. In Nanoscience and technology: a collection of reviews from nature journals (pp. 11-19) (2010).
\bibitem{F26}
R. S. Swathi and K. L. Sebastian, J. Chem. Phys, \textbf{129}, 054703.(2008).
\bibitem{F27}
R. S. Swathi, K. L.Sebastian, J. Chem. Phys, \textbf{180}, 086101. (2009).
\bibitem{F28}
R. S. Swathi, K. L.Sebastian, J. Chem. Phys, \textbf{121}, 777-787 (2009).
\bibitem{F29}
M. Selig, E. Malic, K. J. Ahn, N. Koch, and A. Knorr, Phys.Rev. B \textbf{99}, 035420 (2019).
\bibitem{F30}
H. Zahra, D.Elmaghroui, I.Fezai, S. Jaziri, J. Appl. Phys,  \textbf{120}, 205702 (2016).
\bibitem{F31}
J.Qu, D.Jiang, L.Wang, K.Liu, X.Xu, C.Yao, W.Sun (2019).  Science, \textbf{54}, 8450-8460 (2019).
\bibitem{F32}
K. A.Velizhanin, A. Efimov. Phys. Rev. B\textbf{84}, 085401  (2011).
\bibitem{F33}
 E. H.Hwang, S. D.Sarma , Phys. Rev. B \textbf{75}, 205418 (2007).
\bibitem{F34}
B. Wunsch, T. Stauber, F.Sols, F.Guinea  New Journal of Physics, \textbf{8}, 318.(2006).
\bibitem{F35}
X.Wu,  F.Mu, Y.Wang, H. Zhao, Molecules, \textbf{23}, 2050 (2018).
\bibitem{F36}
F. Tian, J. Lyu, J. Shi, M.Yang . Biosensors and Bioelectronics, \textbf{89}, 123-135 (2017).
\bibitem{F37}
M.Yi,  S. Yang, Z. Peng, C. Liu, J. Li, W. Zhong, R. Yang,  W. Tan , Analytical chemistry, \textbf{86}, 3548-3554.(2014).
\bibitem{F38}
E.M.Narvaez, B.P.Lopez,  L. B.Pires,  A.Merkoci, Carbon, \textbf{50}, 2987-2993 (2012).
\bibitem{F39}
X. Wu, Y. Xing, K. Zeng, K. Huber, and J. X. Zhao,  Langmuir, \textbf{34}, 603-611 (2018).
\bibitem{F40}
P.Zheng, N. Wu, Chemistry–An Asian Journal, \textbf{12}, 2343-2353. (2017).
\bibitem{F41} 
Medintz, I. L., Hildebrandt, N. (Eds.). (2013). FRET-Forster resonance energy transfer: from theory to applications. John Wiley  Sons.
\bibitem{F42}
S.Fishman, D. R. Grempel, and R. E.Prange,  Phys. Rev. Lett.  \textbf{49}, 509 (1982).
\bibitem{F43}
R. Brandenberger, and W.Craig,  The European Physical Journal C, \textbf{72}, 1 (2012).

\bibitem{F44}
P. K. Chow, R. B. Jacobs-Gedrim, J. Gao, T.-M. Lu, B. Yu, H. Terrones, and N. Koratkar, ACS Nano 9, 1520–1527 (2015)
\bibitem{F45}
I. Pelant and J. Valenta, Luminescence Spectroscopy of Semiconductors (Oxford Univ. Press, Verlag, 2012).
\bibitem{F46}
Z. Lin, B. R. Carvalho, E. Kahn, R. Lv, R. Rao, H. Terrones, M. A. Pimenta, and M. Terrones, 2D Mater. 3, 022002 (2016).
\bibitem{F47}
S. Sasaki, Y. Kobayashi, Z. Liu, K. Suenaga, Y. Maniwa, Y. Miyauchi, and Y. Miyata, Appl. Phys. Express 9, 071201 (2016).


\bibitem{F48}
S.Ayari, A.Smiri, A.Hichri, S.Jaziri and T. Amand, Phys. Rev. B, \textbf{98}, 205430 (2018).
\bibitem{F49}
 R.Celotta, and T. B. Lucatorto, (Eds.) Experimental methods in the physical sciences. Academic Press (1999).
\bibitem{F50}
I. L.Medintz, and  N. Hildebrandt (Eds.). (2013). John Wiley and Sons (2013).
\bibitem{F51}
D. Van Tuan, A. M. Jones, M. Yang, X. Xu, and H. Dery, Phys.Rev. Lett. \textbf{122}, 217401 (2019).
\bibitem{F52}
M. Danovich, I. L. Aleiner, N. D. Drummond, and V. I. Falko, IEEE J. Sel. Top. Quantum Electron. \textbf{23}, 6000105 (2016).
\bibitem{F53}
T. Sohier, M. Calandra, and F. Mauri, Phys. Rev. B \textbf{94}, 085415 (2016).
\bibitem{F54}
A. Kormányos, G. Burkard, M. Gmitra, J. Fabian, V. Zólyomi, N. D. Drummond, and V. Falko, 2D Mater. \textbf{2}, 022001 (2015)
\bibitem{F55}
F. A. Rasmussen and K. S. Thygesen, J. Phys. Chem. C  \textbf{119} , 13169 (2015).
\bibitem{F56}
Froehlicher, G., Lorchat, E., and Berciaud, S. (2018). Physical Review X, 8(1), 011007.
\bibitem{F57}
He, J., Kumar, N., Bellus, M. Z., Chiu, H. Y., He, D., Wang, Y., and Zhao, H. (2014). Nature communications, 5(1), 1-5.
\bibitem{F58}
Hill, H. M., Rigosi, A. F., Raja, A., Chernikov, A., Roquelet, C., and Heinz, T. F. (2017). Physical Review B, 96(20), 205401.
\bibitem{F59}
Trushin, M. (2018). Physical Review B, 97(19), 195447.


\end{thebibliography}
\end{document}